\documentclass[prb,twocolumn,aps,showpacs,floatfix]{revtex4-1}

\usepackage[pdftex]{graphicx} 
\usepackage{epstopdf}
\usepackage{verbatim}
\usepackage{color}
\usepackage{subfigure}
\usepackage{tabularx}
\usepackage{amsfonts}
\usepackage{wasysym}
\usepackage{bm}

\newcommand{\be}{\begin{equation}}
\newcommand{\ee}{\end{equation}}

\newcolumntype{C}[1]{>{\centering\let\newline\\\arraybackslash\hspace{0pt}}m{#1}}

\begin{document}
 
\title{Hybrid Weyl Semimetal}

\author{Fei-Ye Li$^{1}$}
\thanks{These two authors contribute equally.}
\author{Xi Luo$^{1}$}
\thanks{These two authors contribute equally.}
\author{Xi Dai$^{2}$}
\author{Yue Yu$^{3,5}$}
\author{Fan Zhang$^{4}$}
\author{Gang Chen$^{3,5}$}
\email{gchen$_$physics@fudan.edu.cn}
\affiliation{${}^1$CAS Key Laboratory of Theoretical Physics, 
Institute of Theoretical Physics, Chinese Academy of Sciences, 
Beijing 100190, People's Republic of China}
\affiliation{${}^2$Beijing National Laboratory for Condensed Matter Physics,
and Institute of Physics, Chinese Academy of Sciences,
Collaborative Innovation Center of Quantum Matter, Beijing 100190, 
People's Republic of China}
\affiliation{${}^3$State Key Laboratory of Surface Physics,
Center for Field Theory and Particle Physics, Department of Physics, 
Fudan University, Shanghai 200433, People's Republic of China}
\affiliation{${}^4$Department of Physics, University of Texas at Dallas, 
Richardson, Texas 75080, USA}
\affiliation{${}^5$Collaborative Innovation Center of Advanced Microstructures, 
Nanjing, 210093, People's Republic of China}
\date{\today}

\begin{abstract}
We construct a tight-binding model realizing one pair 
of Weyl nodes and three distinct Weyl semimetals.   
In the type-I (type-II) Weyl semimetal, both nodes 
belong to type-I (type-II) Weyl nodes. 
In addition, there exists a novel type, dubbed  ``hybrid Weyl semimetal'', 
in which one Weyl node is of type-I while the other is of type-II.
For the hybrid Weyl semimetal, we further demonstrate 
the bulk Fermi surfaces and the topologically protected surface states, 
analyze the unique Landau level structure and quantum oscillation, 
and discuss the material realization.

\end{abstract}

\maketitle

\emph{Introduction.}---Since the theoretical and experimental 
discovery of topological insulator~\cite{KaneReview,ShouchengReview}, 
the study of topological states of matter has become one of the 
major topics in condensed matter physics. Apart from the triumphs 
of systems with full energy gaps, the concept and discovery 
of Weyl semimetals (WSM) have stimulated intensive activities 
in understanding the band topology for gapless systems~\cite{Volovik,Murakami2007,WanXG_PRB,
Burkov2011,AshvinReview,Chen2012,ZhangFanPRL,XiaolingQiReview,
HassanExp,DinghongExp,LuLingExp,BernevigNature,XZZ,WeylMagnon,PhysRevB.86.115133,li2016order}. 
A WSM, in the original setting, has linear conic band crossings 
at the Fermi energy~\cite{WanXG_PRB}. These band crossing points, 
i.e., the ``Weyl nodes'', behave like sources and sinks of the 
Berry curvature in the momentum space and are topologically protected.
Based on the bulk-boundary correspondence, 
the surface state of a WSM takes the form of Fermi arc
that connects a pair of Weyl points with opposite 
chiralities~\cite{WanXG_PRB}. 

\begin{figure}[tp]
\includegraphics[width=0.33\textwidth]{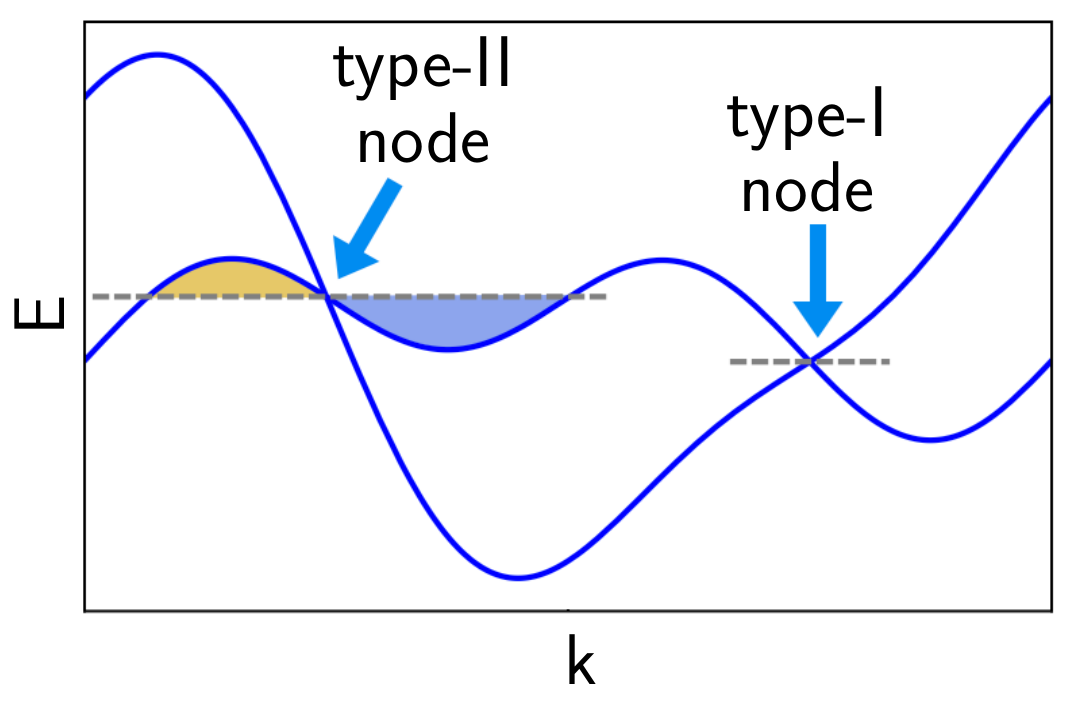}
\caption{(Color online.) A schematic band structure of a 
hybrid WSM with a pair of Weyl nodes. The left (right) node 
is a type-II (type-I) Weyl node. Generically, the energies 
of these two Weyl nodes cannot be identical when both time 
reversal and inversion symmetries are absent. 
}
\label{demo}	
\end{figure}

A novel type of structured Weyl node, dubbed type-II, 
was recently discovered in WTe$_2$~\cite{BernevigNature}
and a spin-orbit-coupled superfluid~\cite{XZZ}. 
In the original WSM, referred as type-I, 
the Fermi surface is composed of discrete Weyl points 
with emergent Lorentz invariance. In type-II WSMs, 
the conic spectrum is tilted near the nodes, and the 
emergent Lorentz invariance is broken. 
These Lorentz-invariance-violating type-II Weyl nodes 
appear at the contact points of the electron and hole pockets in type-II WSMs. 
In all the previous works on type-I or type-II WSMs, 
the two Weyl nodes in a pair with opposite chiralities 
are of the same type~\cite{BernevigNature,Trivedi_arxiv}. 
One may wonder whether it is possible to have a WSM such that 
one Weyl node belongs to type-I whereas its chiral partner 
belongs to type-II (see Fig.~\ref{demo}). 
In this paper, we analyze the band topology of a concrete 
lattice model and demonstrate that the proposed WSM phase 
with mixed types of Weyl nodes can be realized in the concrete model. 
We dub this special type of WSM ``hybrid WSM''.
Remarkably, it is possible to have a {\sl single isolated Weyl fermion}
in the excitation spectrum of this hybrid WSM
rather than several pairs of Weyl fermions in the conventional case. 
We explicitly show that the band structure contains two Weyl nodes,
whose types can be tuned separately and independently. 
Therefore, our model provides a simple platform 
to manipulate the energy-momentum positions, the types of Weyl nodes, 
and the transitions among different types of WSMs. 
We further explore the unique Landau level structure 
and quantum oscillation of the hybrid WSM. Based on our results, 
we propose that the hybrid WSM may be found in magnetically 
ordered non-centrosymmetric materials.

{We start from the classification of the type I and type II Weyl 
nodes following Ref.~\onlinecite{BernevigNature} and Ref.~\onlinecite{XZZ}.
Due to the linear band touching, the original pair of Weyl nodes 
with opposite chiralities has an emergent Lorentz invariance at low energies, 
and the gapless elementary excitation near the nodes 
are often called ``Weyl fermions''. The Lorentz invariance, 
however, is broken by the lattice regularization that necessarily 
connects the two Weyl nodes at high energy~\cite{Franz}.
Significantly, this leads to the intactness of anomalous Hall 
effect but the breakdown of chiral magnetic effect.
More seriously, the violation of Lorentz invariance in 
condensed matter systems allows the tilting of Weyl nodes,
as described in the following general $k\cdot p$ 
Hamiltonian near a Weyl node:
\begin{eqnarray}
h_{\text{Weyl}}({\bm k}) = 
\sum_{ij} k_{i} v_{ij} \sigma_{j} 
+ \sum_i k_{i} u_{i}, 
\label{h3}
\end{eqnarray}
where $i,j=x,y,z$. Interestingly, the sign of $\bar{v}-1$, 
with $\bar{v}_{i}=\sum_jv^{-1}_{ij}u_{j}$, defines the type of Weyl node:
\begin{eqnarray}
&& \bar{v}<1\,\, \Rightarrow \,\, \text{type I} , \label{eqvp}
\\
&&\bar{v}>1\,\, \Rightarrow \,\, \text{type II} . \label{eqvm}
\label{criteria}
\end{eqnarray}
For type I nodes, the tilting is not too strong, and only energy
anisotropy in momentum develops near the node.
For type II nodes, the tilting is sufficiently strong such that 
the Weyl node develops a structure~\cite{XZZ}, 
i.e. a ``bouquet'' of two spheres in mathematics, 
as depicted in Fig.~\ref{demo}. Physically, this implies 
that an electron and a hole Fermi pockets touch at the Weyl node. 
As shown in Ref.~\onlinecite{XZZ}, an isolated ``bouquet'' enjoys 
the same first Chern number of the original Weyl node, 
while the electron or hole pocket is characterized by a zeroth Chern number, 
i.e., the difference in hole-band number across the Fermi sphere.
In general, a Weyl node is characterized by its chirality and its type. 
The chirality cannot be changed by any local perturbation due to 
its topological protection by the unaltered Chern number. 
The type, however, can be modified by local disturbance through 
a topological transition in the zeroth Chern numbers,
which twists the electron (hole) band down (up) near the Weyl node, 
as depicted in Fig.~\ref{demo}. In order to separately manipulate 
the types of the two Weyl nodes with opposite chiralities, 
any symmetry, e.g., inversion or antiunitary particle-hole 
symmetry, that relates the two nodes must then be broken.
This is suggestive of the fundamental existence of a pair 
of hybrid Weyl nodes with opposite chiralities:  
one in type I and the other in type II. 

We here propose a two-band tight-binding model of fermion
hopping on a simple cubic lattice. At low energy this 
minimal model captures the essential physics of one pair 
of Weyl nodes with opposite chiralities. In real 
crystalline solids, it may represent a lattice regularization 
for a WSM or a Weyl superconductor;
in cold atom systems, it may directly describe a Weyl 
superfluid or an artificial optical lattice with Weyl nodes.
Nevertheless, such a Hamiltonian may be written as
\begin{eqnarray}\!\!\!\!\!\!\!\!
H&=&\sum_{j}\big[ {-t_x^{}} {\bm c}_{j}^{\dagger} \sigma_{x}^{} {\bm c}_{j + \hat{x}}^{} 
                 -t_y^{} {\bm c}_{j}^{\dagger} \sigma_{x}^{} {\bm c}_{j+ \hat{y}}^{} 
                 -t_z^{} {\bm c}_{j}^{\dagger} \sigma_{x}^{} {\bm c}_{j+ \hat{z}}^{} 
\nonumber \\
&&               -i t_y' {\bm c}_{j}^{\dagger} \sigma_y^{} {\bm c}_{j+ \hat{y}}^{} 
                 -i t_z' {\bm c}_{j}^{\dagger} \sigma_z^{} {\bm c}_{j+ \hat{z}}^{} 
                 + \mbox{h.c.}\big]
                 + m {\bm c}_{j}^{\dagger} \sigma_{x}^{} {\bm c}_{j}^{}.
\label{h1}
\end{eqnarray}	
Here ${\bm c}^{\dagger}=(c^{\dagger}_{\uparrow},c^{\dagger}_{\downarrow})$ 
are the creation operators of fermions with spin $\uparrow$ and $\downarrow$, 
in which the Pauli matrices $\bm\sigma$ act on; $\bm t$ and $\bm t'$ are 
the hopping energies and $m$ is the onsite energy, which are all spin dependent;
$\hat{x}, \hat{y}, \hat{z}$ are the three first neighbor vectors 
on the cubic lattice. In momentum space, the Hamiltonian Eq.~(\ref{h1}) reads
\begin{eqnarray}
h({\bm k}) = \, && (m-2t_x^{}\cos k_{x}-2t_y^{}\cos k_{y}-2t_z^{}\cos k_{z}) \sigma_{x}  
\nonumber \\
&& + 2t_y'\sin k_{y} \sigma_{y}+2t_z'\sin k_{z} \sigma_{z}.
\label{h2}
\end{eqnarray}

One can easily demonstrate that 
there exists one pair of Weyl nodes 
at ${\bm q}_{\pm}=(\pm k_0,0,0)$ in the bulk Brillouin zone,
and that the Fermi velocities are ${\bm v}_{\pm}=(\pm 2t_x^{}\sin k_0, 
2t_y', 2t_z')$ at the nodes, where $\cos k_0 = (m/2 - t_y - t_z)/t_x$.
One can also check that in this model both time reversal 
and inversion symmetries are broken, as 
\begin{eqnarray}
\mathcal{T}h({\bm k})\mathcal{T}^{-1} &\neq & h({-{\bm k}}),
\\
{\mathcal P}h({\bm k}){\mathcal P}^{-1} & \neq & h({-{\bm k}}), 
\end{eqnarray}
where $\mathcal{T}=\sigma_{y} K$, $\mathcal{P}=I_{2\times 2}$,
$K$ is the complex conjugation, and $I_{2\times 2}$ is an identity matrix. 
Such broken symmetries allow the presence of Weyl nodes, 
but their energies are not necessarily the same.
However, there are emergent inversion-like and antiunitary 
particle-hole symmetries in the model, i.e.,
$\sigma_x h({\bm k}) \sigma_x = h(-{\bm k})$ 
and $\sigma_z h({\bm k}) \sigma_z = -h^*(-{\bm k})$.
The former dictates the two nodes to appear at the same energy.

\begin{figure}[t]
\includegraphics[width=0.36\textwidth]{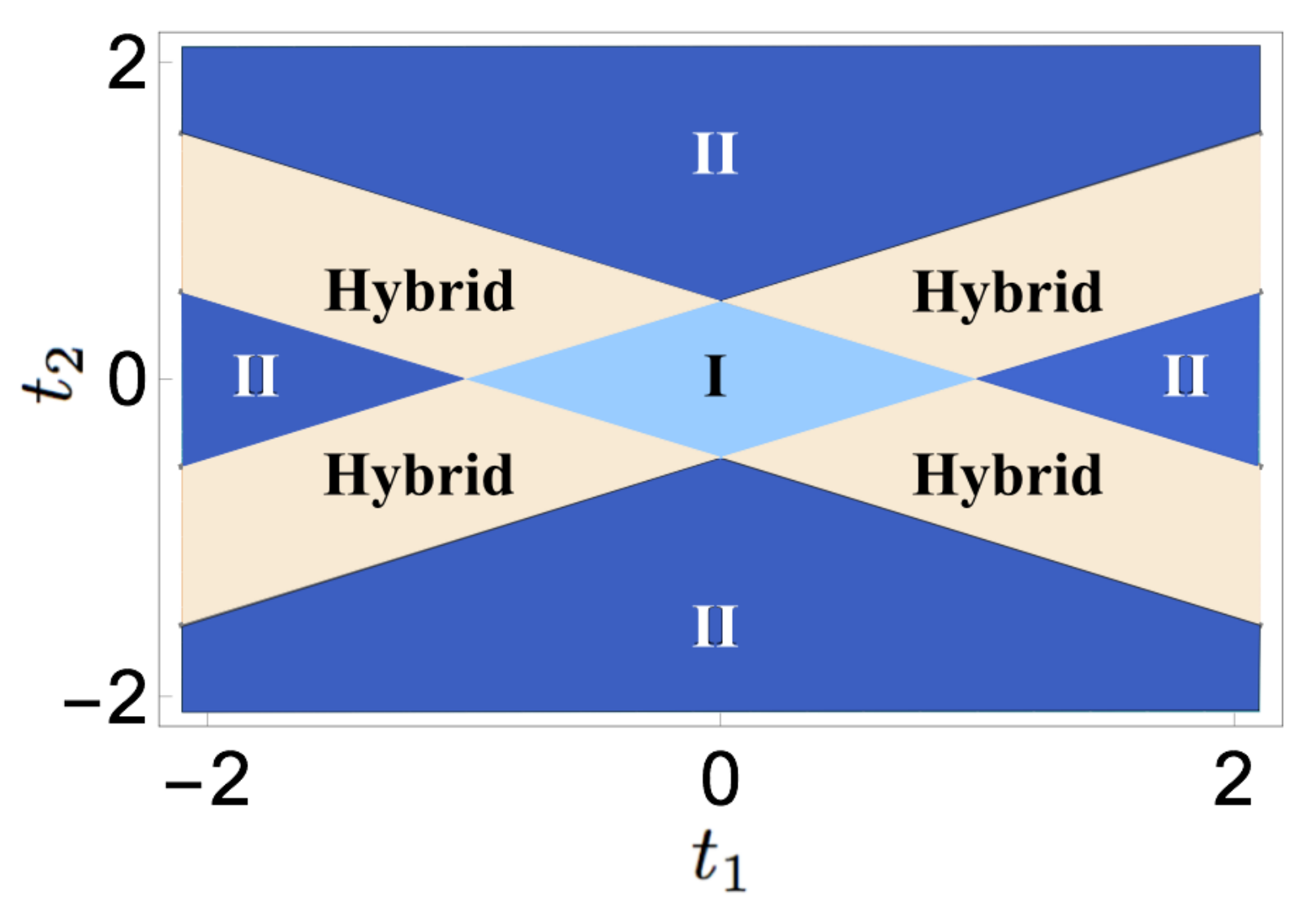}
\caption{(Color online.) The WSM diagram in $t_1$-$t_2$ plane with 
$\phi_1=\pi, \phi_2=\pi/2$. In the light (dark) blue region, 
type-I (type-II) WSM is realized. In the remaining part of the diagram,
hybrid WSM is obtained. See the main text for the detailed discussion.
}
\label{fig2}	
\end{figure}					

The inversion-like and the antiunitary particle-hole symmetries 
both dictate the identical type of the two Weyl nodes. 
Based on the classification criteria, both nodes are of type-I, 
since $\bm{\bar v}_\pm=\bm{u}=0$ in the model. 
In order to convert the type-I Weyl nodes into other types, 
we may introduce additional first and second neighbor hoppings 
into the Hamiltonian as follows:
\begin{eqnarray}
H' = \sum_{j} \big[ t_{1}^{} e^{- i \phi_{1}} 
  {\bm c}_{j}^{\dagger} {\bm c}_{j+ \hat{x}}^{} 
+ t_{2}^{} e^{- i \phi_{2}}  
  {\bm c}_{j}^{\dagger} {\bm c}_{j + 2\hat{x}}^{}
+ \mbox{h.c.} \big],
\label{ht}
\end{eqnarray}
which in momentum space can be expressed as 
\begin{eqnarray}
h'({\bm k}) = 2t_1\cos(\phi_1 - k_{x}) + 2t_2\cos(\phi_2 - 2 k_{x}) .
\end{eqnarray}
Eq.~(\ref{ht}) describes the spin independent hopping processes 
along the ${x}$ direction, and their phase dependence 
lifts the aforementioned inversion-like and antiunitary 
particle-hole symmetries, as long as not both of the two 
phases are integer multiples of $\pi$. Thus, $h'({\bm k})$ 
generically leads to different energy modulations of the 
two original Weyl nodes. Intriguingly, the position and the 
chirality of each Weyl node remain intact, though their 
energies are different.

\begin{figure}
\includegraphics[width=0.48\textwidth]{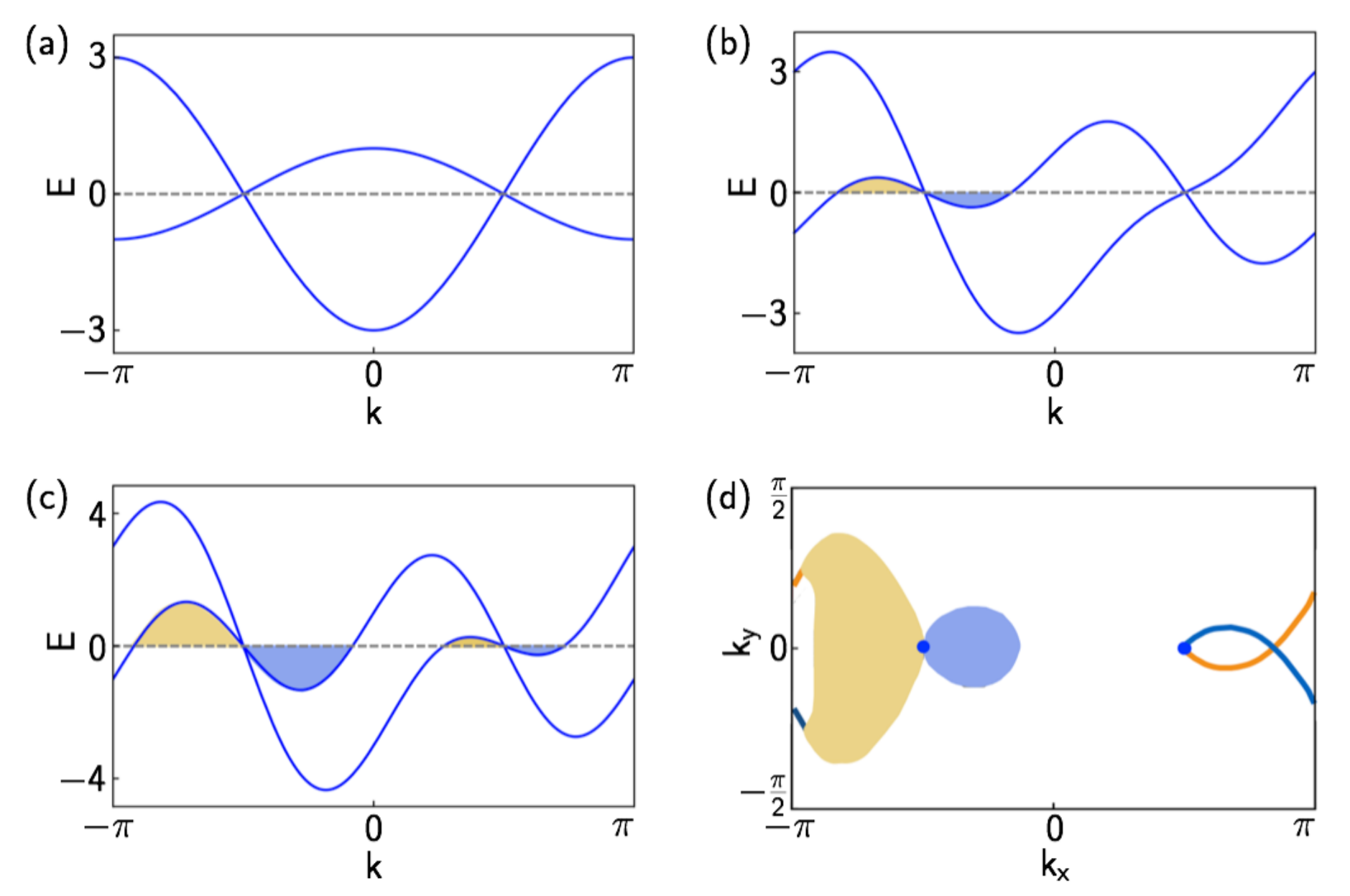}
\caption{(Color online.) The band structure
along $k_x$ direction and the surface states 
of different WSMs with representative 
parameters:
(a) $t_1=0.5,t_2=0$, type-I WSM. 
(b) $t_1=0.5,t_2=0.5$, hybrid WSM. 
(c) $t_1=0.5,t_2=1$, type-II WSM.
The hole pocket (in orange) and the electron pocket 
(in light blue) near the type-II node are indicated.
(d) Surface Fermi arcs of a finite slab along the 
(001) direction for the hybrid WSM in (b). 
Two Weyl nodes (at the blue dots) are projected to 
$(\pm\pi/2,0)$ in the surface Brillouin zone 
($k_x$-$k_y$ plane). The orange (blue) area is the 
projected hole (electron) pocket. The orange (blue) arc is 
localized on the top (bottom) surface and connects the hole 
pocket with the type-I node. The lattice constant is set 
to unity. 
}
\label{dispersion}	
\end{figure}

To examine the types of the deformed Weyl nodes under 
the action of Eq.~(\ref{ht}), we expand the total     
Hamiltonian $H+H'$ near ${\bm q}_{\pm}$ and obtain  
\begin{eqnarray}
h_{\pm}({\bm p}) &=& \mp 2t \sin k_{0} p_{x}\sigma_{x} 
                    + 2t p_{y}\sigma_{y} 
                    + 2t  p_{z}\sigma_{z}
                    \nonumber\\
& + & 2[t_1\sin (\phi_1 \mp k_{0}) 
  +   2t_2\sin (\phi_2 \mp 2k_{0}) ] p_{x},
\label{h4}
\end{eqnarray}
where the subindex ``$\pm$'' refers to the two Weyl nodes at ${\bm q}_{\pm}$
and we have let $-t_x^{}=t'_y=t'_z=t$. To capture the essential physics, 
we set $t=1, t_y=t_z=2, m=8$ throughout the paper.
By tuning other parameters ($t_1,t_2,\phi_1,\phi_2$)
and using the criteria in Eqs.~\ref{eqvp} and \ref{eqvm}, 
we can realize all three WSMs (i) ${v}_{\pm}<1$, (ii) ${v}_{\pm} > 1 $ 
and (iii) ${v}_{+}>1$, ${v}_{-} < 1 $ or vice versa, 
that correspond to type-I WSM, type-II WSM and hybrid WSM, 
respectively. The WSM diagram is depicted in Fig.~\ref{fig2}. 
In Figs.~\ref{dispersion}a,b,c, we further depict 
the band structures of some representative parameters 
for each case. Although we have chosen the parameters 
to make two Weyl nodes occur at the same energy
for hybrid WSM, no symmetry protects such degeneracy 
in the general situation.  Due to the mixed types of 
Weyl nodes in the hybrid WSM, one could actually have 
one single Weyl fermion in the excitation spectrum of the 
hybrid WSM (see Fig.~\ref{dispersion}b). 

The bulk Fermi surface and the surface states of the hybrid WSM 
in Fig.~\ref{dispersion}b are depicted in Fig.~\ref{dispersion}d. 
Rather than directly connecting two Weyl nodes in the type-I WSM, 
the arcs start from the type-I node and terminate when 
overlaping with the projected area of the hole pocket.

	
{\emph{Landau level structure.}---One intriguing property 
of a WSM is its Landau level structure near the Weyl nodes.  
With both type-I and type-II Weyl nodes in its spectrum, 
the hybrid WSM provides a unique opportunity to observe the Landau 
level structures of two distinct nodes in one system. 
For a type-I Weyl node, a simple $k\cdot p$ theory gives Landau 
levels with specific level indices for any direction 
of the magnetic field. The zeroth Landau level is chiral and 
contributes to a non-zero chiral anomaly. By contrast, the Landau 
level structure of a type-II Weyl node depends on the direction 
of the magnetic field~\cite{BernevigNature,PhysRevLett.116.236401,Shengyuan}. 
For the magnetic field along the tilted direction of the type-II Weyl node, 
each Landau level is still characterized by one specific level index and the   
zeroth chiral Landau level (CLL) remains. For the field normal 
to the tilted direction, however, the Landau level 
no longer has a well-defined level index because 
different level indices are now mixed by the kinetic term. 
Despite the absence of the specific Landau level indices, 
the magnetic field cannot open any energy gap at the type-II node. 
The gapless nature at the type-II node is required to 
cancel the non-zero chiral anomaly at the type-I node. 

With the lattice model, we explicitly explore the evolution of  
the Landau level structure as the type of Weyl nodes and the WSMs 
are varied. When the magnetic field is along 
$(100)$ direction, i.e., the tilted direction of the Weyl 
nodes, two CLLs at the two nodes of the type-I WSM remain recognizable 
even for the type-II and the hybrid WSMs (see the left panels in Fig.~\ref{LL}).
By comparison, when the magnetic field is no longer parallel 
to the tilted direction of the Weyl nodes, 
e.g., along the $(110)$ direction, the Landau level structure behaves 
rather differently (see the right panels in Fig.~\ref{LL}). 
The CLL hybridizes with other levels when the corresponding Weyl 
node becomes type-II. Dictated by the chiral anomaly cancellation, 
the type-II Weyl node remains gapless in the hybrid WSM. 
In type-II WSM, however, both two CLLs are not recognizable. 

\begin{figure}[t]
\includegraphics[width=8.7cm]{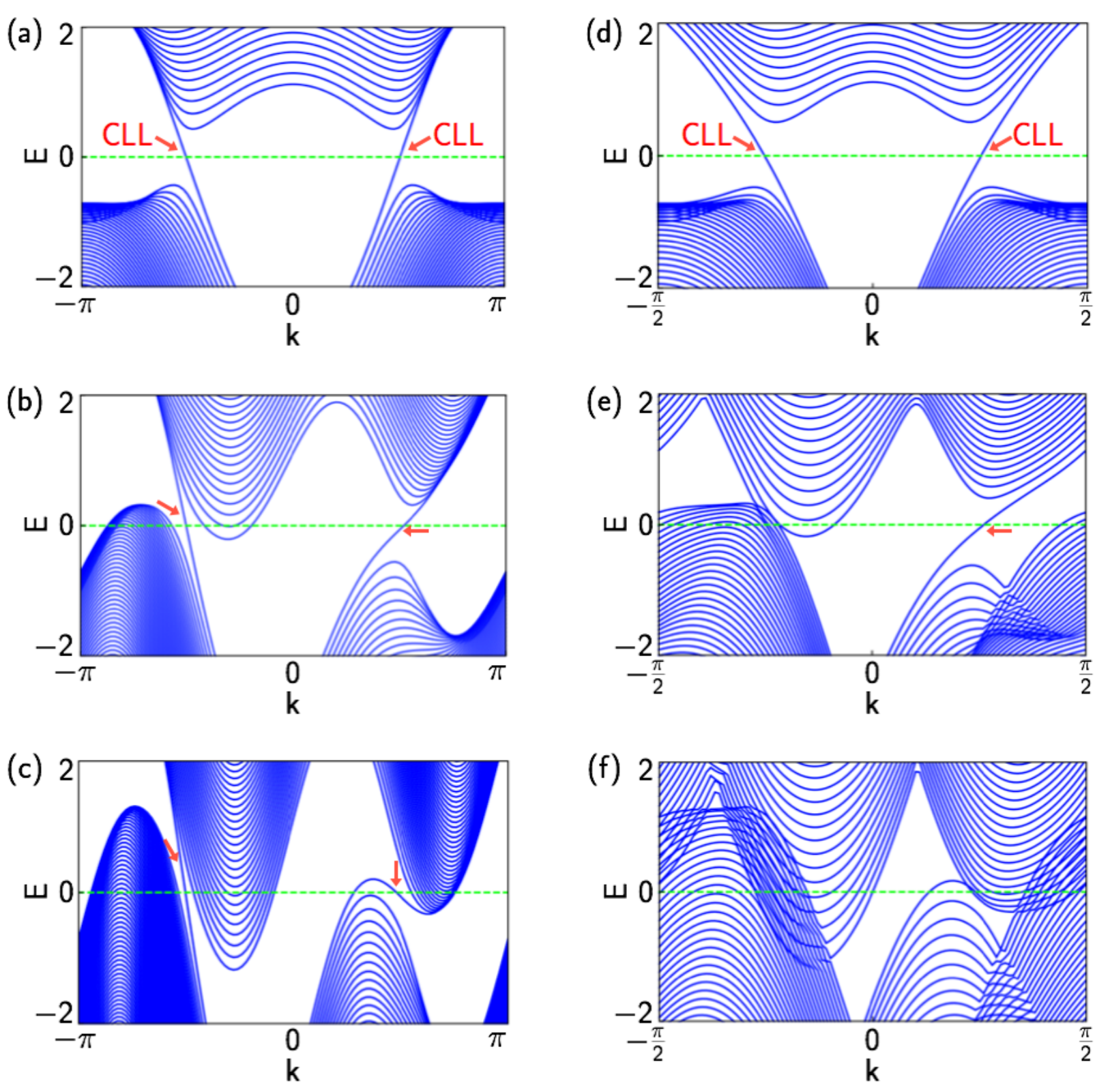}
\caption{(Color online.) The evolution of Landau level structures 
of different WSMs. The (red) arrows indicate the zeroth chiral 
modes near the (dashed) Fermi level. The left panels are the Landau levels 
along the momentum $(k,0,0)$ for field ${\bf B}=B[100],t_1=0.5$ and
(a) $t_2=0$ (type-I WSM);
(b) $t_2=0.5$ (hybrid WSM);
(c) $t_2=1$ (type-II WSM).
The right panels are the Landau levels along the momentum $(k,k,0)$ 
for field ${\bf B}=B[110],t_1=0.5$ and
(d) $t_2=0$ (type-I WSM); 
(e) $t_2=0.5$ (hybrid WSM);
(f) $t_2=1$ (type-II WSM).
We have chosen the Landau gauge
and $B=1/200 \ \Phi_0$ where $\Phi_0$ is the flux quantum.
}
\label{LL}	
\end{figure}

{\emph{Quantum oscillation.}---Now we discuss the quantum oscillation 
of the hybrid WSM. Quantum oscillation directly probes the Fermi surface.
For the type-I WSM with point-like Fermi surface, there is no oscillation 
from the bulk due to the vanishing density of states. It was, however,     
realized that the surface states of a finite slab could contribute to 
quantum oscillation and the period depends on the thickness of the slab~\cite{potter2014quantum}.   
For the hybrid WSM, there are electron and hole pockets at the type-II node. 
Therefore, the bulk quantum oscillation in a hybrid WSM directly detects
the electron and hole pockets contacting at the type-II node. 

Using the recursive Green's function method~\cite{PhysRevB.91.085105}, 
we numerically demonstrate the oscillatory behavior of 
the density of states $\rho(\mu, k_x)$. Here $\rho(\mu, k_x)$ 
is the density of states of the two-dimensional 
sub-system $h({\bm k}) + h'({\bm k})$ with a specified $k_x$ component, 
and $\mu$ is Fermi energy chosen as the energy of the nodes. 
The results are presented in Fig.~\ref{fig5}, for 
the magnetic field along $(100)$ direction in the Landau gauge. 
The oscillation period of $\rho(\mu,k_x)$, 
$\Delta(B^{-1})$, is related to the cross section 
area $S$ of the Fermi surface at $k_x$ with $S={4\pi^2}/{\Delta(B^{-1})}$.
In Fig.~\ref{fig5}, we have chosen two extremal 
cross sections of the electron and the hole pockets.

As we have tuned the Fermi energy at the nodes,
only the electron and hole pockets at the type-II nodes 
contribute to the oscillation periods. Thus, two periods 
are found in Fig.~\ref{fig5}. In the more general case, 
when the Fermi energy slightly deviates from the nodes, 
the electron (or hole) pocket at the type-I node would 
give another period in the 
quantum oscillation, and the total number of 
oscillation periods would be odd. As a comparison, for 
the conventional type-I WSM or type-II WSM where 
two Weyl nodes with opposite chiralities are 
related by certain symmetries, the density of states 
at two Weyl nodes would behave identically, and the 
number of oscillation periods should be even if 
the oscillation periods at each node can be 
separately resolved. 

\emph{Discussion.}---Going beyond the concrete lattice model,   
we ask the necessary conditions to find the hybrid WSMs in physical systems. 
In the search of WSMs, one guiding principle 
is to look for systems that break either time reversal or inversion symmetry~\cite{WanXG_PRB,AshvinReview}. 
This is because with both symmetries 
the Berry curvature vanishes at every point in the momentum space and cannot   
lead to any monopole singularity that was required for a WSM. 
To create monopole singularities of the Berry curvature, 
the necessary condition is to break either time reversal 
or inversion symmetry. The recent discovery of type-I WSMs in the TaAs family
and type-II WSMs in WTe$_2$ both belong to the inversion symmetry 
breaking case~\cite{HassanExp,DinghongExp,BernevigNature}. 
In TaAs and WTe$_2$, the pair of Weyl nodes belongs to the 
same type due to time reversal symmetry. To search for 
hybrid WSM with mixed types of Weyl nodes,  
one should find systems that lack both inversion 
and time reversal symmetries. We here propose that the magnetically ordered 
non-centrosymmetric materials are natural systems
that host hybrid WSMs. Among the existing WSMs, TaAs and WTe$_2$ 
do not have inversion symmetry. To possibly convert them into hybrid WSMs,
one could dope these materials with magnetic ions and create
magnetic orders in them. Alternatively, one could introduce the 
inversion symmetry breaking in the WSMs with magnetic orders.

\begin{figure}[t]
\includegraphics[width=8.7cm]{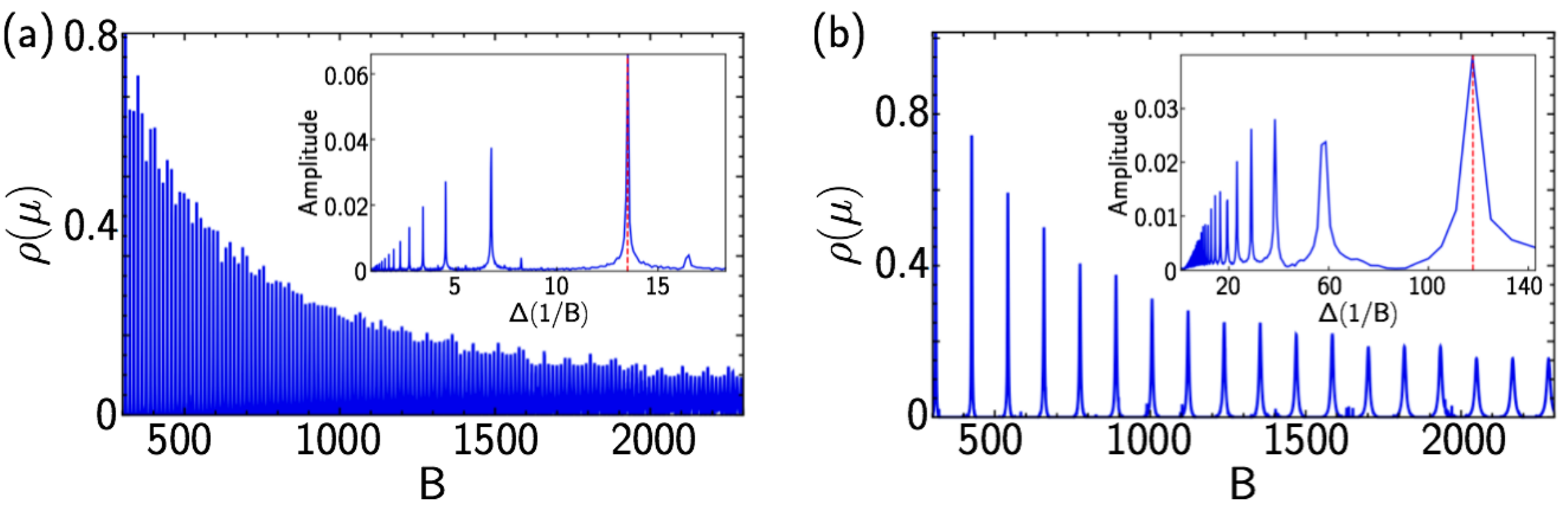}
\caption{(Color online.) 
The oscillation behavior of the density of states 
$\rho(\mu, k_x)$ for the hybrid WSM in Fig.~\ref{dispersion}b.
 ${\bf B}=B(1,0,0)$. 
Fourier spectra in the insets indicate the 
oscillation periods $\Delta(B^{-1})$ (in the red line). 
Other peaks are higher harmonic components, 
$\frac{1}{2}\Delta(B^{-1}),\frac{1}{3}\Delta(B^{-1}), \cdots$.
(a) The extremal cross section 
of the hole pocket occurs at $k_x=-2.4$ and has an area $0.296\pi^2$, 
and $\Delta(B^{-1})=13.5$.
(b) The extremal cross section 
of the electron pocket occurs at $k_x=-1$ and has an area $0.034\pi^2$,
and $\Delta(B^{-1})=117.8$.
In the recursive Green's function method~\cite{PhysRevB.91.085105}, 
the real space degrees of freedom along $z$ direction 
is treated recursively. We set the system size $L_z = 10^5$ 
and choose an imaginary part $\delta=10^{-4}$ 
for the level broadening.
}
\label{fig5}
\end{figure}

To summarize, we present a lattice model to realize three distinct 
WSMs. In addition to the WSMs that were previously known,
we demonstrate the existence of a hybrid WSM with mixed types of 
Weyl nodes. The Landau level structure 
and the quantum oscillation behaviors of the hybrid WSM are discussed. 
We propose possible realizations of hybrid WSM in magnetically 
ordered non-centrosymmetric materials. 

\emph{Acknowledgments.}---We thank Dr.Yang Qi for the positive
comments. This work is supported by the 
973 Program of MOST of China 2012CB821402, NNSF of China 
11174298,11474061 (FYL, XL, YY), NNSF of China, the 973 program 
of China No.2013CB921700, and the ``Strategic Priority 
Research Program (B)'' of the Chinese Academy of Sciences 
No.XDB07020100 (XD), UT Dallas research 
enhancement funds (FZ), and the Start-up fund of Fudan 
University and the National Thousand-Young-Talents 
Program of People's Republic of China (FYL, GC). 

\bibliography{refs}

\end{document}